\documentclass[%
 reprint,
 amsmath,amssymb,
 aps,
]{revtex4-2}

\usepackage{graphicx}
\usepackage{dcolumn}
\usepackage{bm}

\begin{document}

\title{Electronic and Topological Properties of a Topological Insulator Thin Film Sandwiched between Ferromagnetic Insulators}

\author{P. Pigoń$^1$, A. Dyrdał$^2$}
\affiliation{
$^1$AGH University of Krakow, Faculty of Physics and Applied Computer Science, al. Adama Mickiewicza 30, 30-059 Kraków, Poland \\
$^2$Department of Mesoscopic Physics, ISQI, 
Faculty of Physics, Adam Mickiewicz University, ul. Uniwersytetu Pozna\'nskiego 2, 61-614
Pozna\'n, Poland
}

\begin{abstract}
We consider a thin film of a topological insulator (TI) sandwiched between two ferromagnetic (FM) layers. The system is additionally under an external gate voltage.  
The surface electron states of TI are magnetized due to the magnetic proximity effect to the ferromagnetic layers. The magnetization of ferromagnetic layers can be changed by applying an external magnetic field or by varying  thickness of the topological insulator (owing to the interlayer exchange coupling). The change in the magnetic configuration of the system affects the transport properties of the surface electronic states.
Using the Green function formalism, we calculate spin polarization, anomalous Hall effect, and magnetoresistance of the system. We show, among others, that by tuning the gate voltage and magnetizations of the top and bottom FM layers, one can observe the topological transition to the anomalous quantum Hall state.
\end{abstract}

\maketitle


\section{Introduction}
\label{sec:introduction}

The first papers on topological insulators were published in the last century, with the observation of quantum Hall effect~\citep{Klitzing_PRL45p494y1980}. A few years later, Haldane proposed a mathematical model of two--dimensional topological insulator~\citep{Haldane_PRL61p2015y1988}. This model was then extended by  Kane and  Mele~\citep{Kane_PRL95p146802y2005, Kane_PRL95p226801y2005}, who included spin in the model, making a more complete picture of topological insulators. 
These materials act like trivial insulators in bulk, but they are conducting at the edges~\citep{Fruchart_CRP14p779y2013}, where the topologically protected surface states appear~\citep{Qi_PT63p33y2010}. The importance of topological insulators results from their potential application in spintronics devices or in quantum information processing -- mainly because of dissipationless transport and/or possibility to control electron's spin. Nowadays, they are used in electronic and semiconducting devices, e.g., in photodetectors, magnetic devices, and FET transistors~\citep{Tian_M10p814y2017}. A special class of topological insulators is that of magnetic topological insulators, where the quantum anomalous Hall effect can be observed.~\citep{Chang_S340p167y2013}. There are four ways for a topological insulator to become a magnetic topological insulator~\citep{Bernevig_N603p41y2022}. The most popular method is doping the bulk of topological insulators with transition metal elements, e.g., Bi$_2$Te$_3$, Bi$_2$Se$_3$, and Sb$_2$Te$_3$ doped with Cr or Fe~\citep{Yu_S329p61y2010}. But, controlling the distribution and magnetic order of dopants is rather difficult task~\citep{Hou_SA5peaaw1874y2019}. These materials can have many applications, e.g. in information storage and in dissipationless spin and current transport for low-power-consuming electronics~\citep{Chang_S340p167y2013, Lu_PRL111p146802y2013}.

\begin{figure}[ht]
    \centering
    \includegraphics[width=.4\textwidth]{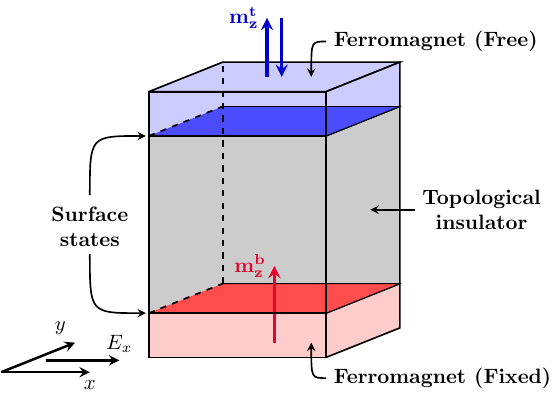}
    \caption{A thin film of a topological insulator sandwiched between two ferromagnets. A red color marks the bottom layer, and the top layer is in blue. The layers are in the $x-y$ plane, while the axis $z$ is normal to the layers. }
    \label{fig:model}
\end{figure}

In this work, we consider a thin film of a topological insulator sandwiched between two ferromagnetic insulators. In such a system, the topological surface states are affected by the magnetic proximity effect. The magnetic layers are magnetized parallel either along or opposite to the $z$--axis (which is normal to the plane of the heterostructure). However, this magnetization orientation can be changed by an external magnetic field.

\begin{figure*}[ht]
    \centering
    \includegraphics[width=0.9\textwidth]{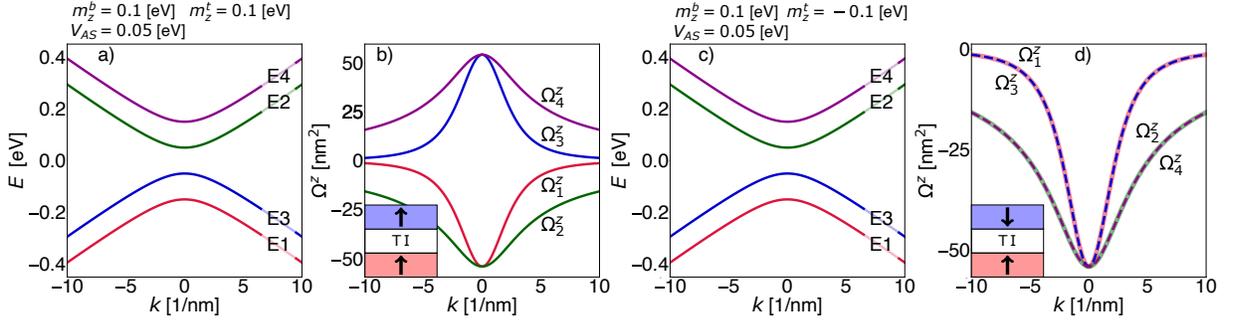}
    \caption{Band structure and the corresponding Berry curvatures for a fixed magnetization of the bottom layer, $m_z^t=0.1$ eV and asymmetric potential $V_{AS}=0.05 eV$. (a) and (b) correspond to the parallel magnetic configuration, i.e., $m_z^t=0.1$ eV, while (c) and (d) to the antiparallel magnetic configuration, i.e., $m_z^t=-0.1$ eV.}
    \label{fig:BC}
\end{figure*}

Here we present the topological and transport properties of such a hybrid structure. In principle we focus on the anomalous Hall effect and on current-induced spin polarization (a.k. Edelstein effect). The anomalous Hall conductivity will give us a direct information about the topological properties of electronic surface states of TI. In turn, the  current-induced spin polarization is one of possible spin-to-charge interconversion effects, that can be responsible for various magnetotransport phenomena and for spin-orbit torque in the considered  systems. 


\section{Model and method}
\label{sec:model}

We consider a thin film of the topological insulator sandwiched between two ferromagnetic insulators with easy axis magnetic anisotropy along the z-direction (Fig.~\ref{fig:model}). The magnetization of the bottom layer is fixed and oriented parallel to the z-axis, whereas the magnetization of the top layer is free to rotate (in an external magnetic field or due to interlayer exchange coupling) from antiparallel to parallel  (or vice-versa) orientation  to the z-axis. 

The Hamiltonian $\hat{H}$ of the electronic surface states of 3D TI, written in the so-called top--bottom representation, takes the form~\citep{Litvinov2020}
%
\begin{eqnarray}
\label{eq:H}
 \hat{H} = v \hat{\tau}_z \otimes (k_y \sigma_x - k_x \sigma_y) + V_{\small{AS}} \hat{\tau}_z \otimes \hat{\sigma}_0 \nonumber\\ 
+(m_{z}^{t} \hat{\tau}_{+} + m_{z}^{b} \hat{\tau}_{-})    \otimes \hat{\sigma}_z , 
\label{eq:}
\end{eqnarray}
where $\hat{\sigma}_i$, $\hat{\tau}_i$ ($i=x,y,z$) are Pauli matrices acting in the spin and layer subspace, respectively, $\tau_{\pm} = \frac{1}{2} (\tau_{0} \pm \tau_{z})$, $\textbf{k} = \{k_x, k_y \}$ is the two-dimensional wave vector, $V_{AS}$ is the asymmetric part of the gate voltage, while $m_z^{t/b}$ denote magnetizations of the top/bottom layer and $\textbf{n}_z=(0,0,1)$.
%
Here  we assume that the layer of TI is thick enough to neglect the effect of hybridization between top and bottom surface states (the effect of hybridization as well as other orientatations of the magnetizations will be published elsewhere).

To characterise the topological properties of the system, we calculate the Berry curvature for each energy band based on the following equation: 
\begin{equation}
  \Omega_n = i \nabla_{\textbf{k}} \times  \left\langle\psi_n\right|\nabla_{\textbf{k}}\left|\psi_n\right\rangle,
\end{equation}
where $\left|\psi_n\right\rangle$ is the  $n$-th eigenstate of the Hamiltonian $\hat{H}$.

Figure~\ref{fig:BC} presents the band structure and the corresponding Berry curvatures for all electronic bands for  two specific cases, i.e., for parallel and antiparalell configurations of the magnetic moments of the two  FM layers (in the following referred to shortly as parallel and antiparallel magnetic configurations). 
In both cases one can see symmetric valence and conduction bands formed by the surface states. Importantly, due to the magnetization and gate voltage, there is a gap in the surface states. For the fixed magnetization of the top and bottom layers, the width of the energy gap can become shrunken, closed, and then reopened depending on the value of $V_{AS}$. Moreover, depending on the magnetic configuration, the sign of the integrated over the angle Berry curvature  changes. In Fig.~\ref{fig:BC} one can see that the Berry curvatures (integrated over the angle) for bands, $E_1$, and $E_3$, are the same in antiparallel magnetic configuration and opposite in the parallel one. This, in turn, leads to the two different topological phases, with Chern number equal $\pm 2$ or $0$, respectively.
The topological phase transition will be seen in the anomalous Hall conductivity, that is directly related to the Berry curvature~\cite{Dyrdal_2017,XiaoRevModPhys2010}:
\begin{equation}
\sigma_{xy} = \frac{e^{2}}{\hbar} \sum_{{\small{n = 1-4}}} \int \frac{d^{2} \mathbf{k}}{(2\pi)^{2}} \Omega_{n} f(E_{n})
\end{equation}
Note that the above expression gives only the intrinsic contribution to the anomalous Hall effect. The scattering effects are neglected here, as we consider quasi-ballistic limit and focuse on the topological properties of the system, i.e., when chemical potential is located inside the energy gap.
\begin{figure*}[ht]
    \centering
    \includegraphics[width=\textwidth]{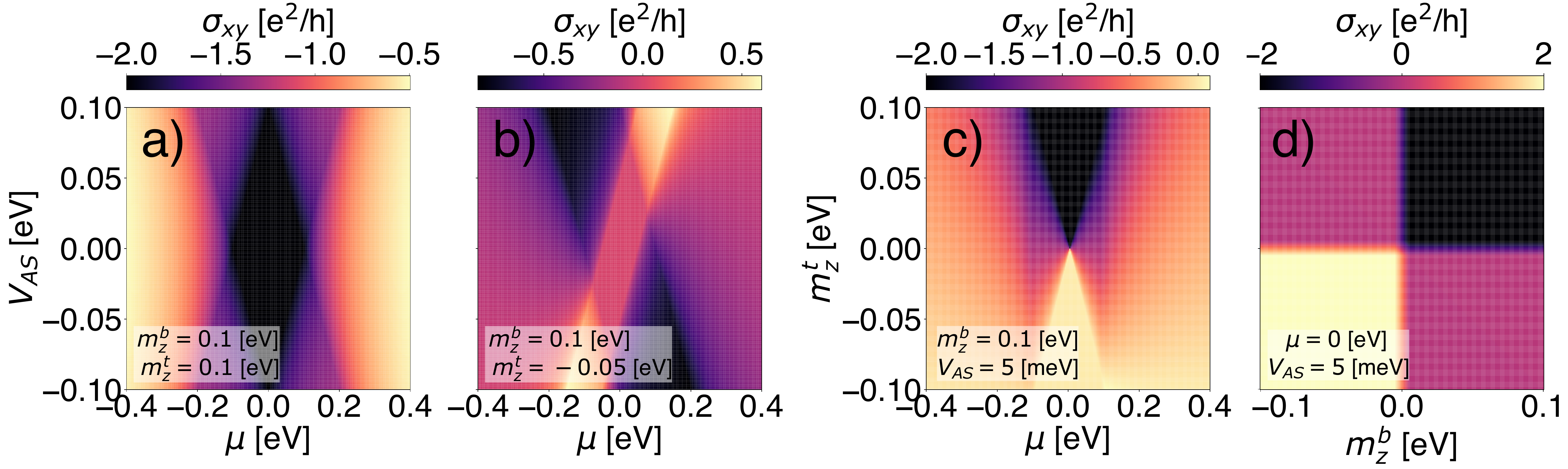}
    \caption{Density plots of the anomalous Hall conductivity as a function of asymmetric potential $V_{AS}$ and chemical potential $\mu$ (a) for parallel magnetization direction ($m_z^t=m_z^b=0.1$ eV) and (b) for parallel magnetization direction ($m_z^t=-0.1$ eV and $m_z^b=0.1$ eV) and as a function of (c) magnetization of the top layer $m_z^t=0.1$ eV and chemical potential $\mu$ with the fixed magnetization of bottom layers $m_z^b=0.1$ eV, $V_{AS}=5$ meV (d) magnetization of the layers with fixed $\mu=0$ and $V_{AS}=5$ meV.}
    \label{fig:DP}
\end{figure*}

In turn, for chemical potential located in valence or conduction bands, we focus on the current-induced spin polarization~\citep{Dyrdal_PRB92p165404y2015,Mahan}
\begin{equation}
    S_{y} = \frac{e\hbar}{2\pi} E_x \int \frac{\mathrm{d}^2 \textbf{k}}{(2\pi)^2} \mathrm{Tr}\left\{\hat{s}_y G_k^R \hat{v}_x G_k^A \right\},
\end{equation}
where $\hat{s}_y = \frac{\hbar}{2} \tau_0 \otimes \sigma_y$ is the spin operator,  $\hat{v}_i=\frac{1}{\hbar} \frac{\partial \hat{H}}{\partial k_i}$ is the velocity operator, and $G_{k}^{R/A}=\left[\left(\mu\pm i\Gamma \right) \hat{\sigma}_0 \otimes \hat{\tau}_0 - \hat{H} \right]^{-1}$ defines the retarded (R) and advanced (A) Green function, respectively, with $\Gamma$ denoting the relaxation rate.
Note that the surfaces of TI are spatially separated. Accordingly, apart from the  spin operator $\hat{s}_y$, that gives the information about the net current-induced spin polarization in the system, one can define the spin polarization induced at a specific surface. The spin polarization at the top/bottom surface is defined by the operator:
\begin{equation}
\hat{s}_{y}^{t/b} = \frac{\hbar}{2} \tau_{\pm} \otimes \sigma_{y}
\end{equation}
and therefore
\begin{equation}
    S_{y}^{t/b} = \frac{e\hbar}{2\pi} E_x \int \frac{\mathrm{d}^2 \textbf{k}}{(2\pi)^2} \mathrm{Tr}\left\{\hat{s}_{y}^{t/b} G_k^R \hat{v}_x G_k^A \right\}.
\end{equation}

\section{Results}
\label{sec:results}

\subsection{Anomalous Hall effect}

Figure~\ref{fig:DP}, presents the anomalous Hall conductivity plotted as a function of the certain parameters defining Hamiltonian~\ref{eq:H}. The most intriguing phenomena happen when the chemical potential lies in the energy gap. The surface states of TI in our model are gapped due to the interplay of gate voltage, $V_{AS}$, and magnetizations of the top and bottom layers, $m_{z}^{t/b}$. 

Figure~\ref{fig:DP}(a) shows  the transverse conductivity as a function  of the chemical potential $\mu$ and the asymmetric potential $V_{AS}$, for fixed values of the magnetization of both layers $m_z^t=m_z^b=0.1$ eV (parallel magnetic configuration). In the middle of the plot, there is a region of the diamond shape, where the conductance  is quantized, $\sigma_{xy}=-2$ $e^2/h$. This region determines the range of the energy gap. With increasing  absolute value of the asymmetric potential, $|V_{AS}|$, this region becomes narrower. In addition, when the chemical potential $\mu$ is far from the bandgap, the Hall conductance diminishes.  
Figure~\ref{fig:DP}(b) shows the conductivity as a function of the chemical potential, $\mu$, and the asymmetric potential, $V_{AS}$.
The magnitude and direction of the magnetization of the top layer are changed in comparison to those in Fig.~\ref{fig:DP}(a), i.e., $m_z^t=-0.05$ eV. This corresponds to antiparallel magnetic configuration, with asymmetric magnitudes of the magnetizations. Now, one can  distinguish four particular regions. The dark areas, where the conductance reaches a maximum negative value (bands $E_1$ and $E_3$ are fully occupied). Thus, by changing the asymmetric potential, one can control the anomalous Hall conductance of the system. However, for this magnetic configuration, there is no area of quantized  conductance.

In turn, in Fig.~\ref{fig:DP}(c) we show the conductivity of the system as a function of the chemical potential, $\mu$, magnetization of the top layer, $m_z^t$, and for the fixed values of the magnetization of the bottom layer $m_z^t=0.1$ eV and for $V_{AS}=5$ meV.  When $m_z^t$ changes from $m_z^t=-0.1$ eV to $m_z^t=0.1$ eV, the magnetic configuration of the system changes from the antiparallel magnetic configuration to the parallel one. Here the changes of amplitude and orientation of $m_{z}^{t}$ determine the width of the energy band gap and the topological phase transition. For $m_z^t<0$ (the antiparallel  magnetic configuration) and chemical potential located in the gap, one observes the trivial phase with zero anomalous Hall conductivity (yellow triangle). In turn, for $m_z^t>0$ (the parallel magnetic configuration) and chemical potential in the gap, the quantum anomalous Hall phase occurs (dark triangle). The possibility of tuning the gap width and topological phase behaviour externally seems to be promising for application in  spintronics and low-energy consumption electronics. 

\begin{figure*}[ht]
    \centering
    \includegraphics[width=\textwidth]{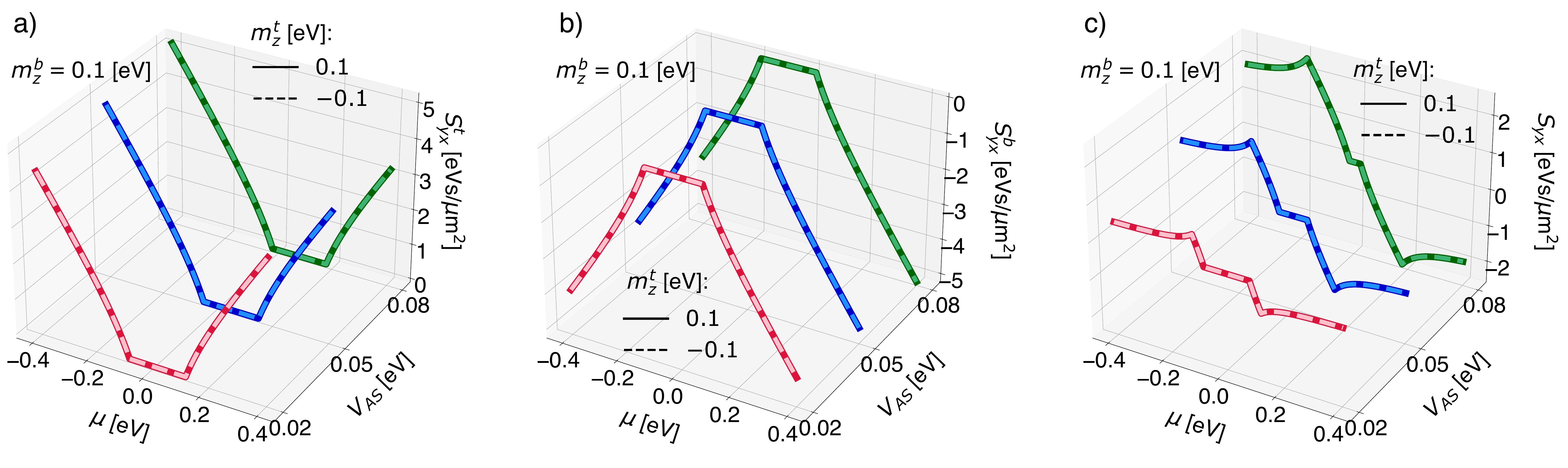}
    \caption{The current induced spin polarization as a function of a chemical potential $\mu$ and asymmetric potential $V_{AS}$ with parallel (solid lines) and antiparallel (dashed lines) direction of magnetization for (a) the top layer, (b) the bottom layer and (c) the total system.}
    \label{fig:CISP}
\end{figure*}

Figure~\ref{fig:DP}(d) presents the variation of the transverse conductance with magnetizations of both layers, for  fixed values of the chemical potential, i.e., $\mu=0$, and $V_{AS}=5$ meV. When the magnetic configuration is parallel, the system is in the quantum anomalous Hall state with  $\sigma_{xy}= \mp 2$ $e^2/h$ (black and yellow areas in Fig~\ref{fig:DP}(d)). The negative sign of anomalous Hall conductance corresponds to the magnetizations along $z$-axis while positive sign of anomalous Hall conductivity refers to the magnetizations oriented antiparallel to the $z$-axis. For the antiparallel configuration of magnetic layers, the system is a trivial insulator state (the purple areas in Fig~\ref{fig:DP}(d)), where the anomalous Hall conductance is zero. 

Based on the above results one can clearly see the characteristic areas where the conductance is quantized. These areas correspond to Chern number equal $\pm2$ depending on whether the magnetic configuration is parallel or antiparallel. Accordingly, when chemical potential lies in the energy gap (fully occupied valence bands) one can observe a topological phase transition to the quantum anomalous Hall insulator state - the state corresponding to the zero longitudinal charge current and fully quantized Hall current. 

\subsection{Current--induced spin polarization}

Now, we consider the  spin-to-charge interconversion effects in the system under consideration, i.e. the spin polarization induced by a charge current. This phenomenon appears as a result of spin-orbit coupling in the system (in the TI layer).  
In Fig.~\ref{fig:CISP}, we present the current-induced  spin polarization due to the electric current flowing along the $x$-axis. The nonequilibrium polarization is shown here as a function of $\mu$ and  $V_{AS}$ for a fixed  magnetization of the bottom layer, $m_z^b=0.1$ eV. Here the orientation of magnetization in the top layer doesn't affect the results (solid and dashed lines overlap in Fig.~\ref{fig:CISP}). Figure~\ref{fig:CISP}(a) shows the spin polarization at the top surface of TI ,$S_{yx}^t$, while Fig.~\ref{fig:CISP}(b) presents the spin polarization at the bottom surface of TI, $S_{yx}^b$. The neat  spin polarization of the system, i.e. sum of the polarizations of top and bottom surface is shown in Fig.~\ref{fig:CISP}(c). 

As only  states that are at the Fermi level contribute to the polarization, and there are no  states at the Fermi level in the energy gap region, then the spin polarization vanishes there, as clearly visible  in the figures for chemical potentials $|\mu|<0.2$ eV. As one can note, the spin polarizations of the topological electron states at the top and bottom layers  have opposite orientations.  Nevertheless, there is some asymmetry between top and bottom polarizations, what can be seen in Fig~\ref{fig:CISP}(a) and ~\ref{fig:CISP}(b). As a result, the net spin polarization of the system is nonzero, as shown explicitly in Fig.~\ref{fig:CISP}(c). 
One should emphasize that pairs of the plots: $S_{yx}^b, \ S_{xy}^t$, $S_{yx}^t,\ S_{xy}^b$ and $S_{yx}, \ S_{xy}$ are its mirror reflection regarding to the line passing through the point $\mu=0$.

\section{Discussion}
\label{sec:summary}

In this work, we considered the electronic and topological properties of the thin film of a topological insulator sandwiched between two layers of a ferromagnetic insulator. Each layer can be characterized by magnetization oriented along the $z$-axis, and we assume that an external magnetic field can control the magnetic configuration of the magnetic layers.
The presented results show that magnetization direction has a significant impact on the transport properties of the system. Firstly, for the parallel orientation of the layer  magnetizations, two of the Berry curvatures are positive, and two are negative. If we switch the magnetization direction of the top layer, four of Berry curvatures are negative. This results in a change of Chern numbers of each band, and results in the change of the behaviour of anomalous Hall conductivity.  Accordingly, depending on the system parameters, $m_{z}^{t/b}$ and $V_AS$, one can switch our system between a trivial insulator state (zero Hall conductance in the energy gap) and a topological insulator state (quantized anomalous conductance in the energy gap).
Moreover, one can induce spin polarization in the system through an external electric field. We showed  that although the chemical potential is in the gap, the current-induced spin polarization appears at each interface and is oriented in a particular direction. 

\section*{Acknowledgement}

This work has been supported by the National Science Centre in Poland under the project Sonata-14 no. 2018/31/D/ST3/02351.
P. Pigoń acknowledges B. Spisak for fruitful discussions. 

\nocite{*}

\bibliography{apssamp}

\end{document}